\documentclass{emulateapj}
\usepackage{graphicx, amsmath, amsthm, amssymb}

\def\ba{\begin{eqnarray}}
\def\ea{\end{eqnarray}}
\newcommand\icarus{{Icarus}}

\shorttitle{Scattering outcomes of  close-in planets}
\shortauthors{Petrovich, Tremaine, \& Rafikov}

\begin{document}

\title{Scattering outcomes of close-in planets: constraints on planet migration}
\author{Cristobal Petrovich\altaffilmark{1}, Scott Tremaine\altaffilmark{2},
and Roman Rafikov\altaffilmark{1} }

\altaffiltext{1}{Department of Astrophysical Sciences, Princeton University, Ivy Lane, Princeton, NJ 08544, USA; cpetrovi@princeton.edu}
\altaffiltext{2}{School of Natural Sciences, Institute for Advanced
  Study, Einstein Drive, Princeton, NJ 08540, USA}

\begin{abstract}
Many exoplanets in close-in orbits are observed to have 
relatively high eccentricities andfangu large stellar obliquities.
We explore the possibility that these result from planet-planet scattering
by studying the dynamical outcomes from 
a large number of orbit integrations in systems with two and three 
gas-giant planets in close-in orbits (0.05 AU$< a < $0.15 AU). 
We find that at these orbital separations, unstable
systems starting with low eccentricities and mutual 
inclinations ($e\lesssim0.1$, $i\lesssim0.1$)
generally lead to planet-planet collisions in which the collision
product is a planet on a low-eccentricity, low-inclination orbit. 
This result is inconsistent with the observations.
 We conclude that
eccentricity and inclination excitation from planet-planet scattering must
precede migration of planets into short-period orbits. This result
constrains theories of planet migration: the semi-major axis must shrink
by 1-2 orders of magnitude without damping the eccentricity and
inclination.
\end{abstract}

\section{Introduction}\label{sec:intro}                                               

Measurements of the orbital eccentricity and 
the projected misalignment between the spin of the host star and 
the orbit of extrasolar planets (i.e., the projected stellar obliquity $\psi$) 
have revealed that planetary
architectures commonly differ from that of our own
Solar System.
The eccentricity distribution is broad, with
$e>0.3$ in $\sim 30\%$ of the planets, while the 
obliquity distribution (mostly from measurements of the 
Rossiter-McLaughlin effect from 
giant planets in close-in orbits)
shows that $\sim 25\%$ of the planets have 
$\psi>30^\circ$\footnote{Taken from
www.exoplanet.org as of November 2013.}.
Similarly, the existence of
massive short-period planets---the so-called Hot 
Jupiters---points towards a dynamical history of known 
extrasolar planets that is quite different from that of 
the Solar System.

Different mechanisms have been proposed to account
for the large eccentricities of giant planets:
interaction with the natal protoplanetary
disk \citep{GS03,OL03},
secular perturbations due to distant stellar \citep{WM03,FT07}
or planetary companions \citep{naoz11,WL11},
dynamical instabilities due to planet-planet gravitational
interactions \citep{LI97,RF96,WM96}, and others.

The mechanisms described above to excite the eccentricities 
that involve gravitational interactions with planetary or stellar 
companions can also potentially account for the observed high stellar 
obliquities and the presence of close-in planets, through tidal 
circularization of high-eccentricity, high-inclination orbits
(e.g., \citealt{FT07,NIB08}).
In contrast, gravitational interactions with the protoplanetary disk 
lead to migration and can account for the planets in close-in orbits,
but are not expected to produce large obliquities
(e.g., \citealt{MN09,KN12}).

Planet-planet gravitational interactions in multiple-planet
systems have been extensively studied through $N$-body
experiments and have been shown to reproduce 
the observed eccentricity distribution of planets 
with periods $>20$ days
reasonably well for a wide range of initial
conditions \citep{FR08,JT08,CFMS2008}.

In this paper, we test a class of planet-formation models in 
which planets either migrate  first or are formed at small orbital 
separations ($a\lesssim0.2$ AU)
  in low-eccentricity and  low-inclination orbits
($e\lesssim0.1$, $i\lesssim0.1$), and then acquire their
eccentricities and inclinations through planet-planet
gravitational interactions. 
This situation could arise, for example, by means of disk migration 
without eccentricity and inclination excitation (or with 
eccentricity and inclination excitation which are 
subsequently damped) followed by disk 
dispersal, or through in-situ formation 
from a close-in disk resulting from a binary merger  
\citep{MST11}.

We recognize that our separation into two phases,  ``migration
in the presence of gas'', and ``planet-planet
gravitational interactions without gas" is artificial
and the planets might enter the gas-free phase
in orbital configurations  that differ from our assumed 
(nearly circular and coplanar) initial conditions
\citep{LT09,LMN13}.
However, unless we make this separation it is hard to isolate
the purely dynamical effects that take place after disk
dispersal.

Two types of instability condition exist for systems containing
a star and two or more orbiting planets.
{\it Hill stability} requires that the planets 
preserve the ordering of the distances to the star,
i.e., there are no orbit-crossing events.
{\it Lagrange  stability} requires both that the planets are Hill stable
and that the outermost planet remains bound to the star 
(e.g., \citealt{BG06}).
For two-planet systems, the former stability criterion has a 
well-defined boundary for the minimum separations of the 
planets with small eccentricities
\citep{gladman93}, while the latter is less 
well-understood
\citep{VM13,DPH13}. 
Note that systems with more than two planets are not provably 
Hill stable for any initial separation
(e.g., \citealt{CWB96,CFMS2008}).

Previous work by \citet{FHR01} showed that the
excitation of eccentricities and inclinations by gravitational interactions 
between planets (``planet-planet scattering'') is largely determined 
by the branching ratios into
different dynamical outcomes such as planetary ejections,
collisions with the star or planet-planet collisions.
In particular, these authors show that  simulations with two 
Jupiter-like planets placed initially in Hill unstable orbits 
result mostly in planet-planet collisions for the range of semi-major
axes $0.5-2$ AU, and that the daughter planets
produced in collisions have small eccentricities ($e<0.05$). 
In contrast, planetary ejections leave the surviving planet 
with significant eccentricity:
 for two equal-mass-planet simulations
\citet{FHR01} find $e\sim 0.4-0.8$ for the survivor.
Similar experiments by \citet{FRY03}  and \citet{FR08} 
show that when the planet masses are unequal 
 the fraction of planet collisions relative 
to ejections is reduced and the eccentricities
of the planets surviving after ejections are lower,
 resulting in a better 
match to the observed eccentricity distribution.

The latter experiments also show that the ratio of ejections 
to planet-planet collisions increases
monotonically with $a/R_p$ in the range 
$(0.5-2) \left(\mbox{AU}/R_J\right)$,
where $a$ is the semi-major axis, $R_p$ is radius of the planet,
and $R_J$ is the radius of Jupiter.
This behavior is related to the ratio $\theta$ of the escape 
velocity at the surface of the planet to the circular velocity of the
 planet:
\ba
\theta^2&\equiv &\left(\frac{2GM_p}{R_p}\right) \left(\frac{a}{GM_\star}\right) \nonumber\\
	&\approx& \left(\frac{M_p}{M_J}\right) \left(\frac{M_\odot}{M_\star}\right)
	 \left(\frac{R_J}{R_p}\right) \left(\frac{a}{0.25 \mbox{ AU}}\right),
	\label{eq:theta}
\ea
for $\theta\gg1$ the planetary radius is so small that 
close encounters mostly lead to ejections relative to collisions, while for 
$\theta <1$ collisions will be more frequent.
Our definition of $\theta^2$ differs from the so-called 
Safronov number because it uses the circular velocity of the 
planet instead of the relative velocity between planets.

Extrapolating the results by \citet{FHR01} or \citet{FR08}  
to smaller orbital separations suggests that
planet-planet collisions should happen even more frequently 
relative to ejections.
However, this conclusion is sensitive to the particular 
initial conditions of these simulations.
In particular, these authors initialize the orbital integrations
with:
\begin{enumerate}
\item small relative orbital separations so the systems are 
Hill unstable,
\item nearly circular and coplanar orbital configurations,
\item and two planets.
\end{enumerate}
With these initial conditions, 
the planets have small relative velocities
(compared to their circular velocities) during
the first close approaches,  so gravitational focusing can significantly 
enhance the rate of collisions.
Additionally, the simulations  by \citet{VM13} 
show that two-planet systems in
Hill stable configurations can still be Lagrange 
unstable over sufficiently long timescales
(at least $\sim 10^5$ orbits
of the inner planet for two Jupiter-mass planets).
Such systems contribute as an extra source of 
ejections and collisions with the star, while avoiding
planet-planet collisions.

Simulations with more than two planets can become Hill 
unstable for larger initial separations between planets than in 
the two-planet case (e.g., \citealt{LI97,LLD98,PT01,MW02}). 
This has the immediate consequence that 
close encounters for these larger separations 
happen after the planets have
attained significant eccentricities, allowing 
close encounters to occur typically at higher relative 
velocities compared to the two-planet simulations,
reducing the overproduction
of collisions due to gravitational focusing.

These simulations show that the timescale  for
instability (time to the first orbit crossing) grows exponentially with 
the initial separation in units of the mutual Hill radii defined in Equation 
(\ref{eq:delta_a})  (e.g., \citealt{CWB96,SL09}).
Once the eccentricities have grown, the number of planets gradually
shrinks due to collisions between planets, planetary ejections,
and collisions with the star, relaxing to a state in which
three or fewer planets persist in eccentric orbits
\citep{JT08}.

We extend previous studies by focusing 
on the dynamical evolution of systems of close-in planets
(0.05 AU$<a<0.15$ AU). We survey the parameter 
space of simulations with two and three planets 
to address whether planet-planet scattering 
can produce eccentricities and inclinations that are as large as 
those of the observed exoplanets.
Recall that our goal is to test the scenario in which 
scattering takes place after the planets 
have migrated to (or formed at) small separations with
small eccentricities and inclinations.

\begin{table*}[ht]
\begin{center}
\caption{Summary of simulated systems and outcomes}
\begin{tabular}{cccccccccccc}
\multicolumn{9}{l}{~~~~~~~~~~~~~~~~~~~~~~~~~~~~~~
~~~~~~~~~~\textsc{Two-planet simulations}}\\
\hline
\hline
Name&K&$ t_{\rm max}$ [yr]&$\sigma_e$$^{\mbox{\tiny{a}}}$&$a$ [AU]&$\beta$&$N_{sys}$  &2 pl.&  C & E  & S &E+S\\
\hline
{\it 2pl-fiducial}  &2& $10^6$&0.01&0.05-0.15 &1&  1,000 & 792 & 184 & 11 &7 &6  \\
{\it 2pl-$K$   }  &3& $10^6$&0.1\mbox{ }&0.05-0.15&1 &   500 & 296 & 118 & 34 & 30&22   \\
{\it 2pl-mass } &2& $10^6$&0.01&0.05-0.15 &1/3&  1,000 & 787  & 194 & 13 & 6 &0  \\
\tableline
\end{tabular}
\begin{tabular}{cccccccccccccccccc}
\multicolumn{11}{l}{~~~~~~~~~~~~~~~~~~~~~~~~~~~~~~~~~~~~~~~~~~~~~
~~~~~~~~~~~~~~~~~~~~~~~\textsc{Three-planet simulations}}\\
\hline
\hline
Name&K&$ t_{\rm max}$ [yr]&$\sigma_x$&$a$ [AU]&$N_{sys}$ &3 pl. &  C & E &S&2C&2E &2S &  C+E&C+S  & E+S & C+E+S &2E+S\\
\hline
{\it 3pl-fiducial}&3& $10^6$&0.01&0.05-0.15       & 1,000 & 255 & 692 & 3      &   0   & 0   &  0 &1 &  21 &  10 &     3 & 15  & 0  \\
{\it 3pl-$K$}&4& $10^7$&0.01& 0.05-0.15                &500 & 83   & 367 & 0      &  0   & 24 & 0 &2 &    12 &  8 &    1  &3    &  0 \\
{\it 3pl-$\sigma$}&3& $10^7$&0.1~~& 0.05-0.15                &500 & 0   & 364 & 6      &  0   & 7 & 2 &0 &    70 &  38 &    0  &13    &  0 \\
{\it 3pl-$a1$}&3& $5\times10^6$&0.01& 0.15-0.45 &500 & 124 & 303 & 13    &  3    & 0   &  0  &3 &   9 & 29 &    11& 6    &  0 \\
{\it 3pl-$a2$}&3& $3\times10^8$&0.01& 2.5-7.5      &500 & 102 & 74   & 158 & 20 & 0   &15 &38 &   2 &   4 &   84 &0    &  3\\
\hline
 \multicolumn{17}{l}{ $^{\mbox{\tiny{a}}}$ $\sigma_i$ is kept constant and equal to 0.01 in all two-planet simulations.}\\
\multicolumn{18}{l}{Note: 2 pl. (3 pl.) means that two (three) planets remain in stable
orbits for a time $ t_{\rm max}$.
C, E, and S
stand for planet-planet collisions,}\\
 \multicolumn{18}{l}{planet ejections ($a>10^3$ AU), and planet collisions  with the star,
respectively. 2C  means that two planet collisions occur, while
 C+E}\\
 \multicolumn{18}{l}{means one collision and one ejection (not necessarily in that order), and so on.}\\
 \multicolumn{18}{l}{
 The stellar mass is Solar and the planet masses are that 
 of Jupiter, except for {\it 2pl-mass} where we use  $\beta=1/3$ and
 $M_1+M_2=2M_J$.}\\
  \multicolumn{16}{l}{  All the planets in the  simulations have radius
  $R_i=R_J$.
}\\
\end{tabular}
\end{center}
\label{table:all_sim}
\end{table*}

\section{Simulations}

We run $N$-body simulations of the evolution of the orbits of giant 
planets orbiting a solar-type star.
In most simulations, the planets are assumed to have 
a Jupiter mass and radius.
Planet-star and planet-planet collisions are assumed
to result in momentum-conserving mergers with no 
fragmentation. 
Collisions are assumed to happen when the
distance between two planets (or planet and star)
becomes less than the sum of their physical radii. 
Since we consider planet-planet collisions, the problem at hand 
is not scale free, i.e., the results depend on the ratio $a/R_p$.

We start by studying simulations with two planets ($N_{pl}=2$) 
because these have been the subject of 
extensive previous work and because understanding this simple case 
facilitates the analysis of simulations with more planets
(e.g., \citealt{FHR01}; \citealt{FR08}).
However, two-planet systems have certain special features that are 
not found in systems with more than two planets. 
In particular, collisions can only happen if the planets initially lie 
inside the well-defined and narrow Hill unstable region 
(see discussion in \S \ref{sec:intro}).
We also perform simulations
with $N_{pl}=3$, which are not subject to the latter
restriction because systems can become Hill unstable
for arbitrarily large initial separations between planets
(e.g., \citealt{MW02}).
 
We do not consider simulations with $N_{pl}>3$.
However, the simulations by
\citet{JT08} with $N_{pl}>3$ almost
always end up with $2-3$ surviving planets after
ejections and collisions of other planets in the system.
Then, as argued by \citet{CFMS2008}, 
 given the chaotic behavior of these systems,  
little memory of the initial number of planets or initial
conditions will remain in the surviving planets.
Therefore, it is plausible that the final state of systems 
initially with $N_{pl}>3$ will be similar to that of systems 
with $N_{pl}=3$.

Even though tidal dissipation can be important for planets that 
approach the host star within a few stellar radii, we ignore its effect. 
We justify this choice because we are interested in studying
the maximum efficiency with which planet-planet scattering
can produce high eccentricities and inclinations, and tides will only 
decrease this efficiency.

\subsection{The code}

We use the publicly available integration 
packages of MERCURY6.2 \citep{chambers99}.
In most simulations, we use MERCURY's Bulirsch-Stoer (BS) 
integration algorithm; we justify this choice because we 
are mostly interested on the evolution of dynamically active 
systems, where planets experience close encounters, and 
the BS algorithm handles close encounters better than 
the other integration algorithms in MERCURY.
We carry out integrations for up to $\sim10^8$ orbits of the inner 
planet with an accuracy parameter $\epsilon=10^{-12}$. 
In one set of simulations ({\it 3pl-K}), we integrate up to $\sim10^9 $ orbits 
of the inner planet using MERCURY's 
Hybrid integrator package, which  integrates orbits symplectically, 
switching to the BS scheme when two planets have a close approach
(defined to be closer than three Hill radii,
as recommended by \citealt{DLL98}).
In {\it 3pl-K} we set the initial time step to be 0.5 days.

\subsection{Initial conditions}

We assume that the initial distributions 
in inclinations and eccentricities follow a Rayleigh law
\ba
dp=\frac{x\,dx}{\sigma_x^2} \exp\left(-\frac{1}{2} x^2/\sigma_x^2\right),
\label{eq:sigma_e}
\ea
where $x=e$ or $i$ and $\sigma_x$ is an input parameter that
is related to the mean, rms, and median eccentricity or inclination by 
$\langle x\rangle=\sqrt{\pi/2}\sigma_x=1.253\sigma_x$, $\langle
x^2\rangle^{1/2}=\sqrt{2}\sigma_x=1.414\sigma_x$,
and $1.177\sigma_x$, respectively.

We measure the inclinations relative to the total angular  
vector of the planets in the initial configuration.

We choose the $\log$ of the semi-major axis to be uniformly distributed 
in a defined range.
Labeling the planets by subscripts $i$ in order of increasing 
semi-major axis, we impose a minimum initial spacing 
of the orbits given by
\ba
\Delta a_{i,i+1}&\equiv& a_{i+1}-a_{i}>K R_{H,i,i+1},\mbox{where} \nonumber \\
R_{H,i,i+1}&=&\left(\frac{M_i+M_{i+1}}{3 M_\star}\right)^{1/3}
\frac{a_i+a_{i+1}}{2},
\label{eq:delta_a}
\ea
and $R_{H,i,i+1}$ is the mutual Hill radius of planets
with masses $M_i$ and $M_{i+1}$.

The initial spacing between orbits mainly changes the 
timescale of onset of dynamical instability
(e.g., \citealt{CWB96}). 
Our choice of $K$ in Equation (\ref{eq:delta_a}) is
empirically guided by the fact that we would like to avoid very 
closely-packed systems that evolve on very short timescales, but still 
have a significant fraction of systems that become unstable
after $\lesssim 10^8$ orbits of the innermost planet.

In most two-planet simulations we use $K=2$.
The Hill stability region for circular orbits 
to the lowest order in the mass ratios 
$M_i/M_\star$ with $i=1,2$ is
$\Delta a_{1,2}>2\sqrt{3}a_1\left[(M_1+M_{2})/ (3M_\star)\right]^{1/3}$
 \citep{gladman93}.
This expression gives $K\simeq3.46$ in Equation (\ref{eq:delta_a})
 but we use a slightly different 
value, $K\simeq3.2$,  because we include the first-order correction 
for Jupiter-mass planets. 
Thus, our choice of $K$ ensures that there is a significant fraction 
of both Hill unstable and stable systems. 

In most three-planet simulations we use $K=3$.
To obtain a crude estimate of the instability time, we rely on 
numerical experiments by  \citet{CFMS2008} using a 
different initial spacing law $\Delta a_{i,i+1}=\tilde{K} R_{H,i,i+1}$ 
(i.e., the spacing is a fixed multiple of the Hill radius, rather than 
exceeding a multiple of the Hill radius). 
These authors show
that for a distribution of planet masses in the
range $(0.4-4)M_J$ the median instability 
timescale\footnote{We follow \citet{CFMS2008}  in using the 
time to the first change in the semi-major axis of any planet  by 
at least $10\%$
as a measure of the dynamical instability growth timescale.} 
can be fitted by the following expression
\ba
\log_{10} (t/\mbox{orbits})=0.021+0.03\exp(1.1~\tilde{K}),
\label{eq:K_tilde}
\ea
where the orbits are those of the innermost planet. 
Note that the instability timescale obtained for a given $\tilde{K}$ 
is a lower limit to that obtained from our spacing law with 
$K=\tilde{K}$ in Equation (\ref{eq:delta_a}).
Thus, by setting $K=3$ the minimum instability timescale 
is $\sim10$ orbits and it rises rapidly to $\sim 10^7$ orbits for 
$\Delta a_{i,i+1}/R_{H,i,i+1}\simeq 5$.

In  \S\ref{sec:K}, we discuss how our results depend on $K$.

\section{Results}

In Table 1, we summarize the input parameters, initial conditions, 
and outcomes of the different simulations.
We have two fiducial simulations, {\it 2pl-fiducial} and 
{\it 3pl-fiducial}, corresponding to two- and three-planet systems.
For these simulations we set the semi-major axis range to 
$0.05-0.15$ AU, $\sigma_x=0.01$, and the maximum integration 
time to $ t_{\rm max}=10^6$ years 
(or $\sim10^8$ orbits for a planet at $a=0.05$ AU).
As discussed in \S2.2, in {\it 2pl-fiducial} we set $K = 2$, 
while in {\it 3pl-fiducial} we use $K = 3$.
In the rest of the simulations, we vary the parameters 
as indicated in Table 1. 
We define the mass ratio for the two-planet simulations
as 
\ba
\beta\equiv\min(M_1,M_2)/\max (M_1,M_2)
\label{eq:beta}
\ea 
and  fix $M_1+M_2=2M_J$. 
 All the planets in the three-planet simulations have Jupiter 
masses, while all planets in the two- and three-planet
simulations have Jupiter radii.

We refer to {\it active} ({\it inactive}) systems as those 
in which the number of planets at the end of the simulation
is less than (equal to) the number at the start. 
The number of planets is reduced by three different 
dynamical events: planet-planet collisions (C), planetary ejections (E),
and planet collisions with the host star (S).

\subsection{Outcomes from two-planet systems}
\label{sec:two_planet}

\begin{figure}[t]
   \centering
  \includegraphics[width=8.5cm]{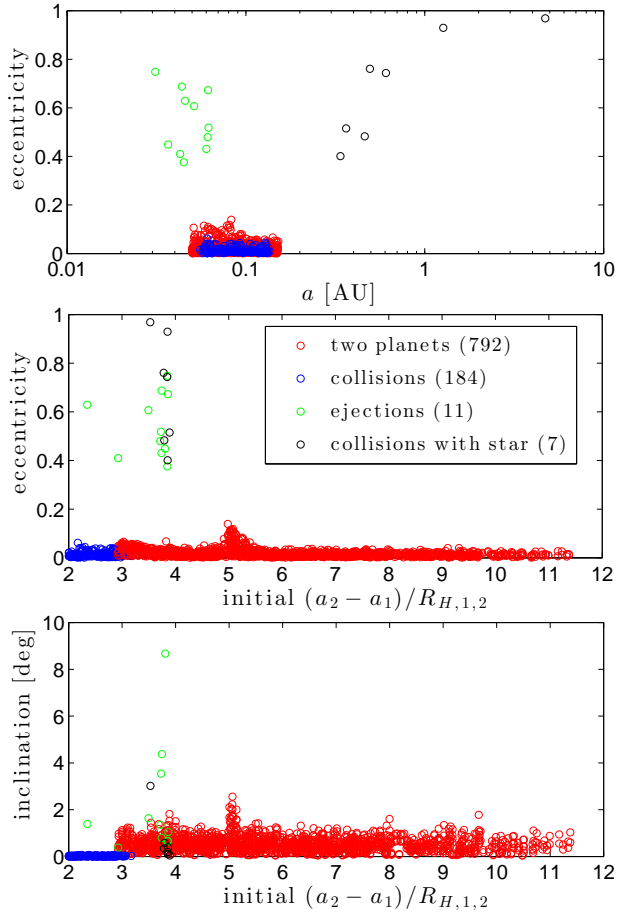} 
  \caption{Final eccentricities and inclinations
as a function of the final semi-major axis (upper panel)
and initial spacing in units of Hill radii (middle and lower panels)
  for our fiducial simulation of two Jupiter-like planets {\it 2pl-fiducial}
  (see Table 1). 
  The colors label the different final outcomes.
Each system was followed for $10^6$ years.
The $\log$ of the initial semi-major axis is drawn uniformly 
distributed in $a=0.05-0.15$ AU with initial spacing 
given by $K=2$ in Equation (\ref{eq:delta_a}).
 The initial eccentricities and inclinations follow a Rayleigh 
 distribution with $\sigma_x=0.01$ (Eq.~\ref{eq:sigma_e}). 
Note that there are two red circles for each system in which both 
planets survive, so such systems might appear more common in 
these scatter plots than they actually are.
Six additional systems are not plotted because one planet escaped 
and the second collided with the star, leaving none behind.}
\label{fig:two_close}
\end{figure}

In Figure \ref{fig:two_close}, we show the results from
our fiducial two-planet simulation {\it 2pl-fiducial}.
The possible outcomes are:
\begin{enumerate}
\item both planets survive on bound orbits until the end of the
integration (``two planets'', red circles)
\item the two planets collide, producing one planet that remains on
a bound orbit  (``collisions'', blue circles)
\item one planet is ejected ($a>1000$ AU) and the other remains on a bound orbit
(``ejections'', green circles)
\item one planet collides with the star, leaving the other on a bound orbit
(``collisions with star'', black circles)
\item one planet collides with the star and the other is ejected, leaving 
no planet in the systems.
\end{enumerate}

From Figure \ref{fig:two_close} and Table 1,
the most common outcome 
is two planets in almost circular and coplanar orbits
($\approx 79 \%$ of the systems). 
The initial separation of these systems is larger than
$\approx 3.2 R_{H,1,2}$, which corresponds to the 
Hill stability boundary  for two
Jupiter-mass planets \citep{gladman93}.
 In these systems the final eccentricities are small (see 
 second panel of Figure \ref{fig:two_close}): 
the mean (median) eccentricity is
0.014 (0.02) and and the eccentricity reaches a maximum of 
 $\approx  0.14$ for a small fraction of orbits with initial spacing of 
 $\approx 5 R_{H,1,2}$ corresponding to the $2:1$ first-order 
 mean-motion resonance. 
The feature seen around the resonance 
agrees with the prediction of \citet{PMT13}
for the time-averaged eccentricity distribution near this 
resonance.

The second most common outcome is planet-planet collisions
($\approx 18 \%$). 
The initial separations of these systems are confined to
 $ \Delta a_{1,2} \lesssim 3 R_{H,1,2}$ and  the collision
 product has a low-eccentricity orbit:
 the mean (median) eccentricity is only 0.014 (0.011).
The small eccentricities of collision products is 
consistent with the results of similar
simulations by \citet{FHR01}, which have a 
median eccentricity of $\approx0.015$. 

The least common outcome is planet ejections and collisions
with the host star, each of which happens only $\approx 1\%$ 
of the time.
The mean (median) eccentricity of the planets left after 
ejections and collision with host star is 0.54 (0.51) 
and 0.68 (0.74), respectively.
Note that most of the ejections and collisions with the star
($\approx89\%$) happen for initial separations 
$\Delta a_{1,2}\gtrsim 3 R_{H,1,2}$ that are Hill stable, but
 become Lagrange unstable after
$\sim 10^5$ orbits of the inner planet \citep{VM13}.

We observe 
that planet-planet collisions happen $\approx11$ and $\approx14$  
times more frequently than ejections and collisions 
with the star, respectively. 
Given that planets in moderate to high eccentricities 
are only produced by planet ejections and collisions 
with the star, we conclude from these simulations that 
excitation of high eccentricities is very inefficient.

The inclinations are always small ($< 10^\circ$) in our simulations 
and the higher values ($\gtrsim 3^\circ$) are reached 
in systems where one planet is ejected. 
The extremely small inclinations (mean of $\simeq0.017^\circ$) in systems with 
collisions are easily understood as a result of angular momentum 
conservation during the collision of two planets.

 We conclude from these experiments that planet-planet scattering 
 in two-planet systems at $a\sim 0.1$ AU is almost never able to 
 generate high eccentricities and inclinations.
This is because the circular orbital velocity at these separations 
is bigger than the escape velocity of the planets 
($\theta^2=0.2$--0.6 from Eq.~\ref{eq:theta}) and, therefore,
planet-planet collisions must happen more frequently than
planet ejections.

\subsection{Outcomes from three-planet systems}
\label{sec:three_planet}

\begin{figure}[t]
   \centering
  \includegraphics[width=8.5cm]{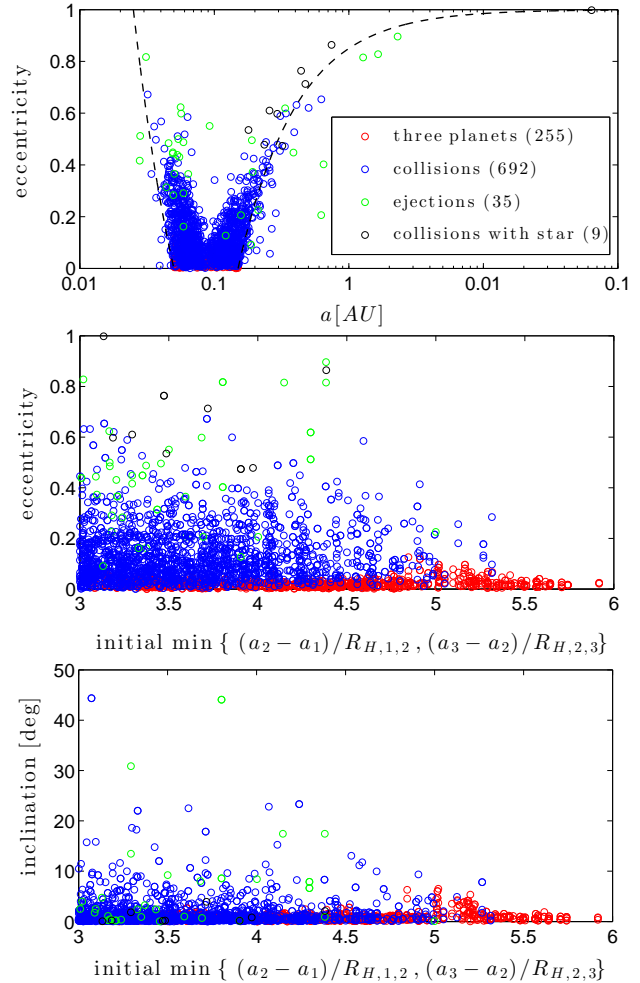} 
  \caption{Final eccentricities and inclinations
as a function of the final semi-major axis (upper panel)
and initial spacing (middle and lower panels)
for our fiducial simulation of three Jupiter-like planets {\it 3pl-fiducial}  
(see Table 1). Each system was followed for $10^6$ yr.
 The colors label the different final outcomes. 
The $\log$ of the initial semi-major axis is drawn uniformly 
distributed in $a=0.05-0.15$ AU with minimum initial spacing 
given by $K=3$ in Equation (\ref{eq:delta_a}).
 The initial eccentricities and inclinations follow a Rayleigh 
 distribution with $\sigma_x=0.01$ (Eq. \ref{eq:sigma_e}). 
 The dashed lines in the upper panel show the curves
 $a(1+e)=0.05$ AU and  $a(1-e)=0.15$ AU.
Note that there are three red circles for each system in which all three
planets survive, so such systems might appear more common in 
these scatter plots than they actually are.
Fifteen additional systems are not plotted because all three
planets were lost.}
\label{fig:three_close}
\end{figure}   

In Figure \ref{fig:three_close}, we show the outcomes from
our fiducial three-planet simulation {\it 3pl-fiducial}.
The different outcomes are summarized in Tables 1 and 2.

We note from Figure \ref{fig:three_close} and Table 1
that most systems ($\approx69\%$) had one planet-planet collision, 
resulting in a final system having two planets.
The second most common outcome ($\approx25\%$) is
a stable system in which all three planets survive for $10^6$ years. 
The third most common channel ($\approx3\%$)
is a system having one collision with either one ejection 
($\approx2\%$) or one collision with the star ($\approx1\%$), 
resulting in a final system with one planet.
Then, we have one planet collision with a subsequent ejection and collision 
with the star ($\approx1\%$), which leaves no planet in the system.
Finally, we have $\lesssim1\%$ systems  with ejections and collisions 
with the star, or a combination of both not involving planet-planet collisions 
 (i.e., E $\cup$  E+S $\cup$  2S in Table 1). 

Other possible outcomes, not seen in {\it 2pl-fiducial} are:
S, 2C, 2E, 2E+S, and 2S+E.

The outcomes 3S, 2C+S, and C+2S  are forbidden by angular momentum
constraints, while  3E, 2C+E, and C+2E are forbidden
by energy constraints.

From Figure \ref{fig:three_close} we see that the contribution to 
the active systems is not strongly confined to any particular initial spacing,
as opposed to the simulations with two planets.
Instead, the ratio of the number of active to inactive systems declines
 (non-red symbols to red symbols)
gradually as the initial spacing increases.
 Equation (\ref{eq:K_tilde}) predicts that
the instability timescale is equal to our integration
time at $\Delta a_{i,i+1}\approx 5 R_{H,i,i+1}$, which
is consistent with the observation that almost all of the systems with 
larger initial separations are inactive.
To address how the initial spacing might affect our results, 
in \S \ref{sec:K} we perform a longer-term set of simulations
with $\Delta a_{i,i+1}> 4 R_{H,i,i+1}$.

\begin{table}[t]
\begin{center}
\caption{Eccentricities and inclinations from the fiducial three-planet simulation}
\begin{tabular}{ccccc}
\hline
\hline
Outcome & 
Number of&
eccentricity &
inc. [deg]\\
&systems&mean-median &mean-median\\
\hline
3 pl&  255   & 0.02 - 0.017 & 0.81 - 0.66 ~ \\
C&  692   & 0.12 - 0.094 & 1.5 - 0.62~ \\
C+E&  21   &0.37 - 0.37 & 1.6 - 0.9  ~~\\
C+S& 9   & 0.67 - 0.61 & 1.0 - 0.2~~ \\
E& 3   & 0.66 - 0.81 & 18 - 11~~ \\
E+S  $\cup$ 2S& 5  & 0.53 - 0.48 & 8.0 - 8.4~~ \\
\hline
Total& 985 & 0.096 - 0.052 & 1.3 - 0.65~ \\
\hline
\end{tabular}
\end{center}
\label{table:three_pl}
\end{table}

The ratio of the number of planet-planet collisions to 
the number of ejections and collisions with the star 
increases from $\approx 6$ in the two-planet simulation
to $\approx10$ in the three-planet case.
The prevalence of collisions over ejections or collisions
with the star is consistent with similar three-planet 
simulations by \citet{johansen12}.

Recall that in the two-planet simulations the ejections
and collisions with the star happen predominantly for an initial spacing 
larger than $\Delta a_{1,2}\gtrsim 3.2R_{H,1,2}$, while planet-planet collisions 
are all confined to $\Delta a_{1,2}<3.2R_{H,1,2}$. 
In contrast, in three-planet systems the different outcomes in 
active systems are not confined to any particular range of initial 
spacing (middle panel of Figure \ref{fig:three_close}).

 From  Table 2 the mean (median) eccentricity
 for systems that had one planet-planet collision and no other 
events (category C; blue circles in Figure \ref{fig:three_close})   
is $ 0.12$ ($0.094$).
These values are larger than the mean eccentricity of 0.014 that
results from collisions in the two-planet simulations.
The category C systems have a characteristic 
fork-like eccentricity distribution.
This distribution roughly agrees with the condition that the collision product
has to pass through a region that was initially populated by planets:
the interval between $a(1-e)$ and $a(1+e)$ must intersect 
$[0.05,0.15]$ AU, as indicated with dashed lines in the upper panel
of Figure \ref{fig:three_close}.

High eccentricities ($e>0.4$) are mainly
reached in systems that had either ejections or collisions 
with the star, but the fraction of these systems is too small ($\lesssim 4\%$)
 to make an important contribution to the eccentricity distribution.
By restricting the sample to the active systems,
 we find that $77\%$ ($92\%$) of the planets have $e<0.2$ ($e<0.3$),
 while all the planets in inactive systems have $e<0.13$.
 
The inclinations are very low in most systems:
$\simeq 95\%$ and $\simeq99\%$  of the systems have $i<5^\circ$
and $i<10^\circ$, respectively.
From Table 2, the mean and median inclinations  are 
$1.3^\circ$ and $0.65^\circ$,  respectively.
The low inclinations arise because most of the active systems 
($\approx 70\%$) experience planet-planet collisions,
which produce  
low-inclination collision products as a result of angular
momentum conservation during the collision of  two planets with
nearly the same initial orbital planes.

In conclusion, the three-planet simulations confirm the main result 
found in the two-planet case:  planet-planet scattering from nearly 
circular and coplanar orbits at small semi-major axes almost never 
produces high eccentricities and inclinations, because planet 
collisions are the dominant 
outcome and these tend to produce merger products with nearly
circular and coplanar orbits.

\begin{figure}[t!]
   \centering
  \includegraphics[width=8.5cm]{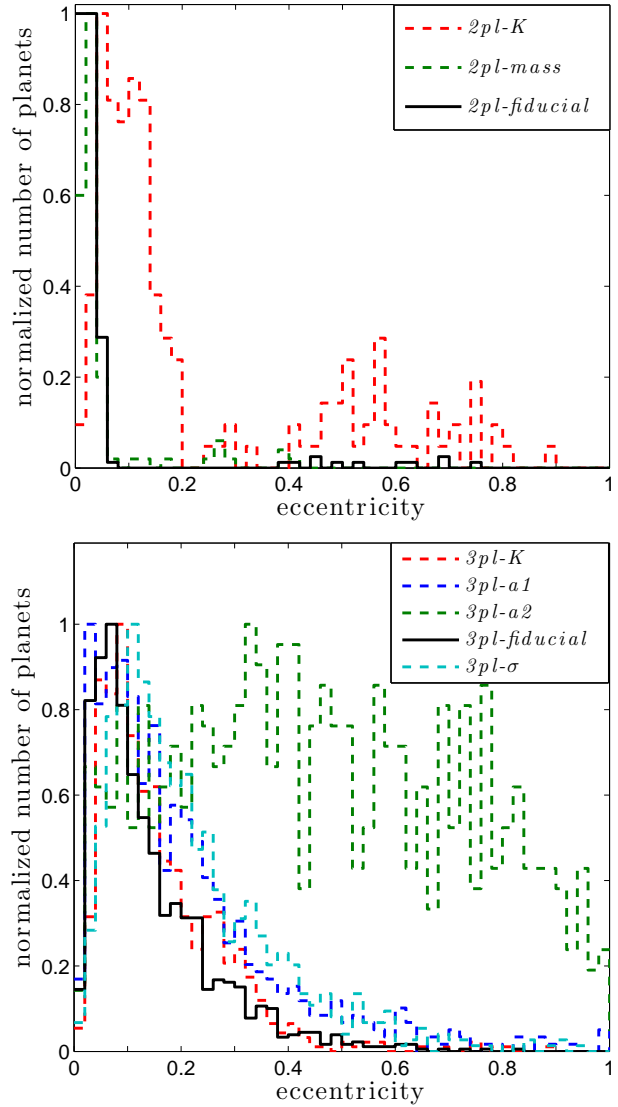} 
  \caption{
 Final eccentricity distribution of all planets in the active systems
  for the simulations with two (upper panel)
  and three planets (lower panel).
The labels indicate the name of the simulation (see Table 1).
All histograms are normalized so the tallest bin has unit height.
  The bin width is 0.02.}
\label{fig:inputs}
\end{figure}

\subsection{Eccentricity distribution}
\label{sec:ecc_coll}

In Figure \ref{fig:inputs}, we show the final eccentricity distribution 
 for all the surviving planets in active systems in our fiducial two- 
 and three-planet simulations 
 (solid black lines in upper and lower panels, respectively). 
We ignore inactive systems because their relative contribution 
depends strongly on the initial conditions and integration time of the 
simulation.
These systems, however, have even smaller eccentricities and 
inclinations and, therefore, make even stronger our conclusion 
that the production of high eccentricities and inclinations 
is inefficient.

The fiducial two-planet simulation has a strong peak 
at $e\lesssim0.06$, which corresponds to the systems that had 
collisions, while the few systems that had either ejections or 
collisions with the star  populate the distribution at $e>0.35$.
For the fiducial three-planet simulation, the distribution is strongly peaked 
at $e\lesssim0.1$ due to the systems that had planet-planet 
collisions and is wider than in the two-planet case with collisions 
contributing to the distribution up to $e\sim0.6$.
The handful of planets with $e>0.6$ are the result 
of ejections or collisions with the star.

In the two-planet simulations, collisions happen very early on
in the simulation with $\approx98\%$ of these events taking place 
during the first $10^3$ years.
Before the collision planets experience
an average of $1.5$ close encounters 
(instances in which the relative distance between two planets
is less than one Hill radius, not counting the last encounter in which 
planets merge).
This suggests that collisions occur before repeated close 
encounters can excite the eccentricities.
 In this case planets would normally collide with low eccentricities, 
 which would produce collision products in low eccentricity 
 orbits\footnote{For planets colliding with low eccentricities, one can show 
 that the mass-weighted eccentricity vector is conserved during the collision:
${\bf e}_c= (1-\nu) {\bf e}_1 + \nu {\bf e}_2, \mbox{where }
\nu=M_2/(M_1+M_2)$ (e.g., \citealt{HP86}).}.
Indeed, the typical eccentricity $e\lesssim 0.05$ seen in the 
collision products is consistent with analytical estimates of the 
eccentricities of collision products based on the assumption of low 
eccentricities for the pre-collision orbits 
 \citep{FHR01,GS03}.

In contrast, in the three-planet simulations the collisions 
are more uniformly distributed in time.
For instance, $\sim 30\%$ ($\sim 20\%$) of 
the collisions happen after $10^3$ years ($10^4$ years).
The number of close encounters before a collision is also larger 
in the three-planet simulations (average of 6.7 compared 
to 1.5 for the two-planet simulations), suggesting that the 
eccentricities of the pre-collision orbits are higher and thus that 
the eccentricity of the collision product should be larger, as observed.
Consistently, the number of close encounters is correlated 
with the eccentricity after collisions (the correlation coefficient
is $\approx0.49$) and almost all ($\approx93\%$) of the
planets with $e>0.4$ had more than 10 close encounters
before the collision.

\begin{table*}[t]
\begin{center}
\caption{Mean and median eccentricities of simulated 
systems and sample of observed exoplanets.}
\begin{tabular}{cccccc}
\hline
\hline
Simulation & 
ecc. active&
ecc. inactive &
ecc. all &
ecc. observed\\
&mean - median  &mean - median &mean - median &mean-median\\
\hline
{\it 2pl-fiducial}  & 0.067 - 0.012   & 0.019 - 0.015 &       0.024 - 0.014    & 0.19 - 0.17\\
{\it 2pl-$K$}       &  ~0.27 - 0.12~   &  ~0.10~ - 0.11~~~& 0.15 - 0.10  &0.19 - 0.17\\
{\it 2pl-mass}  &  0.034 - 0.015   &0.020 - 0.015&    0.022 - 0.015     &0.19 - 0.17\\
{\it 3pl-fiducial} & ~0.13 - 0.098 & 0.020 - 0.017 &  0.096 - 0.052     &0.19 - 0.17\\
{\it 3pl-$K$}      &  ~0.15 - 0.11~   & 0.017 - 0.015  & ~ 0.12 - 0.085    &0.19 - 0.17\\
{\it 3pl-$\sigma$}      &  ~0.21 - 0.16~   & ---  & ~0.21 - 0.16~    &0.19 - 0.17\\
{\it 3pl-$a1$}    &  ~0.19 - 0.14~   & 0.021 - 0.018 &   ~0.13 - 0.061  &0.29 - 0.18\\
{\it 3pl-$a2$}    & ~0.45 - 0.44~ & 0.023 - 0.018&   ~ 0.31 - 0.25 ~       &0.27 - 0.23\\
\hline
\hline
 \multicolumn{5}{l}{Note: we select a sample of exoplanets from 
 www.exoplanet.org as of November}\\
 \multicolumn{5}{l}{2013 in which the planets have $M \sin(i)>0.1M_J$
 and semi-major axis within the }\\
  \multicolumn{5}{l}{range of each simulation. We exclude the planets with $a<0.07$ AU
  to avoid the}\\
  \multicolumn{5}{l}{excess of planets with small eccentricities likely due to tidal circularization.}\\
    \multicolumn{5}{l}{For reference, the number of inactive systems is given 
    in Table 2 in the columns}\\
   \multicolumn{5}{l}{2 pl. and 3 pl. for the two- and three-planet simulations.
   Note that {\it 3pl-$\sigma$}  has no}\\
      \multicolumn{5}{l}{inactive systems.}\\
\end{tabular}
\end{center}
\label{table:ecc_all}	
\end{table*}

\section{Dependence on input parameters}
\label{sec:input}
We make an exploratory analysis of how our results
depend on the properties of the two- and three-planet
systems. 
Recall that the setup of each simulation and the outcomes 
are summarized in Table 1.

In Figure \ref{fig:inputs}, we show the final
eccentricity distribution for each simulation
considering all the remaining planets in active systems.
In Table 3, we show the mean and median eccentricities
for the active, inactive, and all (active + inactive) systems in each simulation
and the same for a sample of observed exoplanets
in the same semi-major axis range.

\subsection{Initial orbital spacing: $K$}
\label{sec:K}

We first test how our results depend on the initial spacing
by running $500$ two-planet simulations with $K=3$,
compared to $K=2$ in the fiducial two-planet simulation (see 
description and outcomes of {\it 2pl-K} in Table 1).
This choice of $K$ means that we exclude the initial spacings that give rise 
to most planet-planet collisions (see upper panel in Figure \ref{fig:two_close}).
In order to have a significant number of unstable systems
we increase the initial eccentricities by setting $\sigma_e=0.1$
in Equation (\ref{eq:sigma_e}), which would expand the 
Hill stability boundary by an approximate factor of 
$\sim (1+8/27(e_1^2+e_2^2)/0.01)^{1/2}$, where $e_1,e_2\lesssim 0.1$
are the initial eccentricities \citep{gladman93}. 
This set-up has no direct theoretical motivation, but it
might be regarded as the result of a previous dynamically active 
phase in the history of the system.

From Table 1, the ratio of the number of collisions to the number of 
ejections and collisions with the star decreases
from $\approx 6$ in the fiducial two-planet simulation to $\approx 1.1$.
This result is expected from the larger initial eccentricities 
in {\it 2pl-K}, which reduces the merger cross section.
From Figure  \ref{fig:inputs} and Table 3, we see that the eccentricity 
distribution broadens and the median eccentricity grows from 0.012 
in the fiducial simulation to 0.12.
This behavior is the result of two effects:
the higher number of ejections and collisions with the star relative 
to planet-planet collisions
(systems with $e>0.2$ in Figure \ref{fig:inputs})
and higher eccentricities after collisions
(systems with $e<0.2$).
The latter effect is expected from the higher initial eccentricities 
and the larger number of close encounters before collisions  
(the average increases from 1.5 in the fiducial simulation to 12), because 
both effects promote higher eccentricities and, therefore, 
collisions at higher relative velocities.

Despite these differences, the main conclusion of this paper still holds: 
planet-planet scattering cannot efficiently excite eccentricities and 
inclinations at small semi-major axes. 
Although the mean and median eccentricities of 0.27 and 0.12 of the 
active systems in {\it 2pl-K} bracket the analogous quantities in the 
observed exoplanet sample (0.19 and 0.17), (i) the initial conditions 
of the simulation already had mean and median eccentricities of 
0.125 and 0.117 so only part of the final eccentricity is due to excitation; 
(ii) realistic ensembles would contain many inactive systems, in 
which almost no eccentricity excitation occurs; (iii) in systems 
with $K=3$ and zero initial eccentricity almost all of the systems 
are inactive.

The mean (median) inclination in active systems is only 
$0.83^\circ$ ($0.04^\circ$).
We note that {\it 2pl-K}  starts with a wide eccentricity 
distribution (mean eccentricity of 0.125) close to the 
dispersion-dominated regime 
($|{\bf e}_1-{\bf e}_2|\gtrsim[(M_1+M_2)/M_\star]^{1/3}=0.126$, where
${\bf e}_i$ is the eccentricity vector of planet $i=1,2$) 
and small inclinations well in
the shear-dominated regime
(relative inclinations $\ll[(M_1+M_2)/M_\star]^{1/3}$).
Such systems can excite eccentricities much more efficiently than the 
inclinations \citep{IM92,RS10},
which is consistent with the simulations
where the final inclinations are much smaller than the final 
eccentricities.


We carry out a similar experiment for the three-planet systems, 
increasing $K$ from 3 to 4 and increasing the integration time from
$10^6$ to $10^7$ years (see simulation {\it 3pl-K} in Table 1).
We integrate 500 systems, using the Hybrid algorithm to speed up the 
integrations as described in \S2.1.


From Table 1, the ratio of the number of collisions 
to the number of ejections and collisions with the star is
$\approx13$, similar to that from the fiducial 
three-planet simulation of $\approx10$. 
This result suggests that this ratio is independent of $K$
 in the three-planet systems.
 This conclusion is not unexpected since the planet-planet 
 collisions in the three-planet simulations come from a wide 
 range of initial separations 
 (middle panel of Figure \ref{fig:three_close}).
 
An outcome that was not present in the fiducial 
simulation is two successive planet-planet collisions 
(2C in Table 1) leaving a single planet with $3M_J$.
This outcome has a branching ratio of $ 2\%$ in the simulation 
{\it 3pl-K}, and appears to require a minimum orbital spacing
of $\approx4.5R_{H,i,i+1}$, where the fiducial simulation contains 
only a small fraction of active systems because of
its shorter integration time. 

From Figure \ref{fig:inputs}, the eccentricity distribution in {\it 3pl-K} 
is slightly broader than the fiducial three-planet simulation.
From Table 3, the small increase in mean and median eccentricity 
still falls short compared to the observed exoplanets.
Likewise, the mean (median) inclination is  only $2.0^\circ$ 
($1.0^\circ$).

 In summary, the effect of increasing the initial spacing is to 
 increase the final eccentricity of the active systems, while 
 decreasing the fraction of active systems. The combined effect 
 of these two changes is not strong enough to alter our conclusion
that eccentricity and inclination excitation is inefficient at small 
semi-major axes.

\subsection{Initial semi-major axis: $a$}
\label{sec:a}

From Equation (\ref{eq:theta}),  we observe that for a fixed 
planet to star mass ratio the physical parameter that governs
 the relative branching ratios for ejections (or collisions with
 the star) and collisions is $a/R_p$. 
Given that most giant planets have radii
close to $R_J$, one expects that the branching ratios
are mostly determined by the semi-major axis.

In the simulations {\it 3pl-a1} and {\it 3pl-a2}, we place three planets 
with the $\log$ of the semi-major axis uniformly
distributed in the range  $a=0.15-0.45$ AU and 
$a=2.5-7.5$ AU, which are larger than the semi-major axis 
range in our fiducial simulation by factors of 3 and 50.
 
We first describe our results for {\it 3pl-a2}, the simulation 
with the largest semi-major axes.
From Table 1, the ratio of the number of collisions 
to the number of ejections and collisions with the star decreases
dramatically from  $\approx10$ in the fiducial three-planet simulation 
to  $\approx0.17$. 
This sharp decrease is expected since the planets in {\it 3pl-a2} have
 $\theta^2=10-30$ compared to $\theta^2=0.2-0.6$ in the fiducial 
 simulation.

From Figure \ref{fig:inputs},  the eccentricity distribution 
in active systems broadens significantly relative to the fiducial
three-planet simulation. 
The median eccentricity of the active systems rises from 
0.098 in the fiducial simulation to 0.44 in {\it 3pl-a2} while the 
median eccentricity of all the systems is 0.25, compared to the 
observed median eccentricity of 0.23 in this semi-major axis range.

Likewise, the distribution of inclinations in  {\it 3pl-a2} broadens 
significantly and the mean (median) inclination in active systems
becomes $15^\circ$  ($10^\circ$).

We next turn to the simulation {\it 3pl-a1} in which the planets 
have $\theta^2=0.6-1.8$.
From Table 1, the ratio of the number of collisions 
to the number of ejections and collisions with the star decreases
from $\approx10$ in the fiducial three-planet simulation 
to $\approx5.6$. 
This reduction in the relative number of collisions broadens the 
eccentricity distribution (Figure \ref{fig:inputs}): the median
eccentricity of the active systems increases from $0.098$ in the 
fiducial simulation to $0.14$ in {\it 3pl-a1}.
These higher  eccentricities in {\it 3pl-a1} are, however, 
still smaller than the measured eccentricities
(Table 3).
Finally, the mean (median) inclination is $3.2^\circ$ 
($1.0^\circ$), slightly larger than the fiducial three-planet
simulation (Table 2).

In conclusion, placing the planets in larger semi-major axis
enhances the efficiency to produce ejections and collisions with the star 
and, therefore, the outcomes of active systems have
higher eccentricity and inclination orbits. 
Our simulation {\it 3pl-a2}  shows that scattering at $a=2.5-7.5$  AU 
produces eccentricities in the active systems that are larger than 
those measured from the observed exoplanets in this semi-major 
axis range, so the results could be consistent with the observations 
if a significant fraction of the observed systems are inactive.
In contrast, the simulations with planets at $a<0.45$ AU 
({\it 3pl-fiducial} and {\it 3pl-a1}) predict eccentricities that are smaller 
than those seen in our exoplanet sample, 
even if we restrict ourselves to active systems, which is mainly 
due to the prevalence of planet collisions.
We remark that the evolution of the different branching ratios
as a function of $a/R_p$
has been previously studied by \citet{FHR01} in simulations
with two planets and our results point in the same direction:
the number of collisions per ejections decreases
with increasing $a/R_p$.

\subsection{Planet mass ratio: $\beta$}
\label{sec:nu}

As shown by \citet{FRY03} and \citet{FR08}, 
the ratio of collisions to ejections decreases as the mass ratio 
$\beta$ in Equation (\ref{eq:beta})  shrinks.

To test how sensitive our results are to the mass ratio,
 we ran 1,000 simulations similar to the fiducial two-planet
simulation, changing the mass ratio from $\beta=1$ to
$\beta=1/3$ (simulation {\it 2pl-mass} in Table 1).
We initialize the system with two planets with masses 
$0.5M_J$ and $1.5M_J$ and randomly pick one planet
to have the smaller semi-major axis.
We fix the planets' radii to that of Jupiter\footnote{This mass-radius
relation is also equivalent to fixing the planets' escape velocity to that of
Jupiter (i.e., $M/R=M_J/R_J$) because it results in the 
same condition for collisions for this particular value of the 
mass ratio $\beta$: distance between planets $<2R_J$.}.

From Table 1,  the ratio of the number of systems that have planet 
collisions to the number of systems with ejections and
collisions with the star increases from 6 for the fiducial simulation
to 10 for {\it 2pl-mass}, although this change is not significant given 
the Poisson errors in the simulations.
We do not see the reduction in the
number of collisions observed in the 
simulations by \citet{FRY03}. 
This difference might be due to the fact that most of the ejections 
and collisions with the star happen in systems that are Hill 
stable, which are excluded in the simulations by \citet{FRY03}.

If a planet is ejected, it is always the lighter one, 
while the median eccentricity of the planet left is 0.26.
This median eccentricity falls within the range
$\sim0.2-0.4$ found by \citet{FR08} for the same mass ratio and
is lower than that observed in the fiducial 
two-equal-mass-planet simulation.

From Table 3, the mean and median eccentricities of
{\it 2pl-mass} are smaller than those of the observed exoplanets, 
as is the case for the fiducial two-planet simulation.
The mean (median) inclinations are $0.57^\circ$ ($1.1^\circ$).

In summary, the overall effect of having unequal-mass 
planets is to produce lower eccentricities after
ejections (at least for $\beta=1/3$). 
Changing $\beta$ to more extreme values would not
change our main conclusion because, even if the number
of collisions per ejection decreases, \citet{FR08} showed that
the eccentricity of the surviving planet after the ejection
would decrease with decreasing $\beta$
(for reference \citealt{FR08} find $e\lesssim0.2$ for $\beta=1/9$).
Likewise, the final inclinations after ejections are also limited 
as one decreases $\beta$ (e.g., \citealt{timpe13}).
In conclusion, the excitation of eccentricities and inclinations is 
inefficient even if the mass ratio departs from unity.

\subsection{Initial eccentricity and inclination: $\sigma_x$}
\label{sec:sigma}

In {\it 3pl-$\sigma$} we initialized the orbits with a broader
eccentricity and inclination distribution 
($\sigma_i=\sigma_e=0.1$ in Eq. [\ref{eq:sigma_e}];  mean of 0.125)
compared to our fiducial three-planet simulation 
($\sigma_i=\sigma_e=0.01$).

From Table 1, the ratio of the number of collisions 
to the number of ejections and collisions with the star 
decreases from $\approx 13$ in the fiducial three-planet 
simulation to $\approx 3.2$.
Similar to the simulation {\it 2pl-K} discussed in \S\ref{sec:K},
this is expected from the larger initial eccentricities and 
inclinations in {\it 3pl-$\sigma$}, which reduces the merger cross 
section.
From Figure  \ref{fig:inputs} and Table 3, we see that the eccentricity 
distribution in active systems broadens and the median eccentricity 
grows from 0.098 in the fiducial simulation to 0.16.

The mean and median eccentricities of 0.21 and 0.16 of the 
active systems in {\it 3pl-$\sigma$} bracket the analogous 
quantities in the observed exoplanet sample (0.19 and 0.17).
However, as discussed in \S\ref{sec:K} for {\it 2pl-K}, 
only part of the final eccentricity is due to excitation because 
the initial conditions of the simulation already had mean and 
median eccentricities of 0.125 and 0.117.
In addition, realistic ensembles would contain many 
inactive systems, in which almost no eccentricity excitation occurs.
thus, our conclusion that planet- planet scattering cannot 
efficiently excite eccentricities and inclinations at small 
semi-major axes should still hold.

The distribution of inclinations in {\it 3pl-$\sigma$}  also broadens 
significantly relative to the fiducial simulation 
and the mean (median) inclination in active systems
becomes $6^\circ$  ($4.5^\circ$). 

\section{Discussion}
\label{sec:discussion}

The main result from our orbit integrations is simple:
at small semi-major axes,  gravitational interactions between planets
in unstable systems mostly lead to collisions rather than to 
excitation of highly eccentric and inclined planetary orbits.

Collisions are more common than ejections or collisions with the star
 because the orbital velocities at 
small separations are typically larger than the escape velocity
of giant planets (see Eq.~\ref{eq:theta}, for reference).
A similar conclusion is reached in the numerical
study by \citet{johansen12}.
Moreover, planet-planet collisions from low-eccentricity, 
low-inclination orbits typically produce collision products that 
occupy orbits with even lower eccentricity and inclination.
The low-inclination orbits are easily understood as a result of 
angular momentum conservation during the collision of two planets 
with nearly the same initial orbital planes.

Consequently, our simulations with two or three planets initially in
low-eccentricity orbits at $a=0.05-0.15$ AU (i.e., excluding {\it 2pl-K}
and {\it 3pl-$\sigma$}) 
predict mean (median) final eccentricities for the active systems that are
 $\simeq1.3-8$ ($\simeq1.5-14$) times smaller than those observed in the 
 exoplanets in a similar semi-major axis range (Table 3). 
Moreover, if we include planets from the inactive systems the mean 
and median eccentricities will be even smaller.
For instance, the mean (median) eccentricities for all (active + inactive)
 planets in each simulation is $\simeq1.6-9$ ($\simeq2-12$) 
 times smaller than the observed values.

The previous result holds even for the simulations with three 
Jupiter-like planets initially at semi-major axes as large as 
$a=0.15-0.45$ AU because the mean (median) eccentricities of 
planets in active systems is 1.5 (1.3) times smaller than
those observed in the data. 
This result breaks down when placing the planets at 
$a=2.5-7.5$ AU because these simulations predict mean 
and median eccentricities in active systems that are larger by a factor of 
$\simeq 2$ than those observed in the data. 
However, this simulation can be made consistent with the data 
by adding a significant fraction of inactive systems.
The eccentricities in the simulations above are larger that those
in the fiducial simulation because of the dramatic decrease in the 
relative number of planet collisions in active systems as we 
increase the semi-major axis (or $\theta^2$ in Eq.~\ref{eq:theta}).

In {\it 2pl-K} and {\it 3pl-$\sigma$} we initialized the orbits with a 
broad eccentricity distribution (mean of 0.125) that might result from
 a previous dynamically active phase when the gas disk is present
\citep{LT09,LMN13}.
As discussed in \S\ref{sec:K} and \S\ref{sec:sigma}, these
simulations starts close to the dispersion-dominated regime, 
in contrast to the rest of the simulations which are well in the 
shear-dominated regime \citep{HP86}.
We observe, however, that even in these more extreme conditions 
planet-planet collisions dominate the branching ratios, limiting the 
eccentricity excitation. 
By considering that realistic ensembles would contain a
mixture of active and in actives systems our conclusion that
scattering at small semi-major axis
is unable to explain the observations can still hold.

Our simulations suggest that in order to 
have final eccentricities that are as high as in the observations the 
initial eccentricities had to be previously excited to $e\gtrsim0.1$.
In principle, eccentricity excitation could be produced by 
effects that we have ignored in our simulations.
For instance, a combination of tides with the host star, secular
precession effects, and
secular interactions with a relatively distant companion 
can lead to eccentricity excitation of a close-in planet
\citep{WG02,correia12}.
However, this mechanism requires a companion in particular
orbital configurations and specific values of tidal dissipation 
that might not be present in the bulk of planetary 
systems \citep{correia12}.

Another possibility is that during the assembly of our 
initial systems, the planets had 
crossed high-order mean motion resonances 
(e.g., $3:1$, $4:1$, and so on)  that could lead to eccentricity and 
inclination excitation \citep{LT09}. 
This possibility is speculative and would require convergent
migration of planets in which case planet-planet 
scattering is likely to have occurred before the planets reach
semi-major axis that are as small  as in our simulations.
Also, the gravitational interactions with the protoplanetary
disk could potentially excite eccentricities during 
the assembly of our initial conditions \citep{GS03,tsang14}.

 We expect that simulations with more than three planets
 would present a similar behavior as our three-planet 
 experiments. Our expectation is guided by the experiments
 of \citet{JT08} who find that regardless of the initial conditions
 and initial number of planets, the systems almost always 
 reach a point in which there are three planets.
There might be two shortcomings with this reasoning.
First, having many planets in active systems can leave the three
planets in orbits with significant eccentricities, which reduces
the merger cross section.
Second,  after several merger events the planets can increase
their masses and, therefore, their escape velocities.
Both effect can reduce the rate of collisions.

Tidal dissipation with the host star should significantly affect 
the long-term dynamical evolution of close-in planets.
\citet{MTR08} suggest that indeed some of the observed 
eccentric close-in planets are in the process of getting 
tidally circularized.
We have, however, ignored the effect of tides in our calculations.
We expect  that the main effect would be to damp the 
eccentricities and given that we are mostly interested in determining 
the maximum efficiency of  planet-planet scattering
 at producing high eccentricities, 
the effect of tides would only make our conclusion stronger.

\subsection{Planet-planet scattering and migration}

From our main result, we can draw the following 
straightforward conclusion:
if planet-planet scattering is responsible for the highly 
eccentric and inclined planetary orbits seen in many exoplanets 
with small semi-major axes, then the scattering must have occurred at 
much larger orbital separations.

Interestingly, this conclusion links the migration 
mechanism or assembly process with the excitation of 
eccentricities and inclinations.
For instance, we can rule out the scenario in which planets 
undergo disk-driven migration to small radii with no
eccentricity and inclination excitation.
We can also rule out the scenario in which planets
are assembled in close-in nearly circular and planar
orbits, for instance as a result of a binary merger,
as proposed by \citet{MST11}.

Our results would support a scenario in which 
planet-planet scattering excites the 
eccentricities and inclinations
either before or early on in the migration process, 
so long as these are not damped during migration.

In this context, \citet{MA05} did a parametric study of type 
II migration coupled with planet-planet scattering
and found that in order to explain the observations
disk migration requires that $|\dot{e}/e| \lesssim 3~|\dot{a}/a|$,
where $\dot{e}<0$ ($\dot{a}<0$) is the time-derivative of the eccentricity
(semi-major axis).
For eccentricity damping more efficient than this limit, 
 planets end-up with eccentricities that are too small by the 
 time they reach  $a\sim 0.1$ AU.
On the other hand, observational constraints from planet pairs
thought to be trapped in orbital resonances during convergent
migration \citep{LP02,GS13}, as well as hydrodynamic 
simulations\footnote{Note, however, that hydrodynamic simulations 
show that for massive planets ($M>5 M_J$) disk-planet interaction can 
lead to eccentricity excitation (e.g., \citealt{Bitsch13,DAA13}).} 
(e.g., \citealt{CDKN07,Bitsch11,KN12})
seem to indicate that $|\dot{e}/e|\gg |\dot{a}/a|$.
This would mean that eccentricity damping occurs much faster than
migration, so even if scattering excites eccentricities early on during
migration these eccentricities would be damped by the time the 
planet migrated to $a\sim 0.1$ AU.  
This might pose a problem for disk-driven migration scenarios.
Direct hydrodynamic simulations coupled with the gravitational 
interaction between planets \citep{MRA08,MA12,LMN13} or planet population 
models \citep{ILN13} might shed light on this issue.

These difficulties with disk migration are not present (or, at least, replaced 
with different difficulties) in models of the formation of Hot Jupiters 
through high-eccentricity migration. 
In these models, planets are scattered to high eccentricities and 
inclinations at large semi-major axes, and then migrate inwards 
through tidal dissipation by the host star. 
Numerical simulations have shown 
that this mechanism can 
produce close-in  planets in highly eccentric and inclined orbits
(e.g., \citealt{NIB08,BN12}).

\section{Summary}

We carried out a large number of orbit integrations of 
close-in two- and three-planet systems (0.05 AU$<a<0.15$ AU) 
and studied whether planet-planet scattering
can excite high-eccentricity and high-inclination orbits.  
We show that this mechanism is almost 
never able to excite eccentricities larger than 
$\sim 0.3$ and inclinations larger than $10^\circ$, and  
the most likely outcome  is typically $e\lesssim0.1$ and 
$i\lesssim5^0$. 

This remarkably low efficiency of eccentricity and inclination
excitation results from the prevalence of planet-planet collisions,
which is an unavoidable outcome of unstable systems in regimes 
where the circular velocity exceeds the escape velocity from 
the planets.

We conclude that if planet-planet 
scattering is responsible for the highly eccentric and inclined 
planetary orbits seen in many exoplanets with small semi-major 
axes, then the scattering must have occurred at 
much larger orbital separations.
This result constrains theories of planet formation by 
disk migration: the eccentricities and inclinations observed in 
extrasolar planets must be excited before or early in the 
migration process, and migration must shrink the semi-major axis 
by $\sim 1-2$ orders of magnitude without damping the 
eccentricity or inclination.

\acknowledgements 

 CP acknowledges support from the 
CONICYT Bicentennial  Becas Chile fellowship. 
ST thanks Henk Spruit for the conversations that stimulated 
this paper, and the Max Planck Institute for Astrophysics, the 
Alexander von Humboldt Foundation, and the Miller Institute for 
Basic Research in Science, University of California Berkeley for 
support and hospitality.
All simulations were carried out using computers 
supported by the Princeton Institute of Computational 
Science and Engineering.



\end{document}